***A Bayesian Framework for Uncertainty-Aware Estimation of Main Pulmonary Artery Velocity Profiles from Phase-Contrast MRI***

Amirreza Kachabi[1] & Naomi C. Chesler[1*]

1) Edwards Lifesciences Foundation Cardiovascular Innovation and Research Center, Department of Biomedical Engineering, University of California, Irvine, Irvine, CA, USA

**ORCIDs of authors:**

Amirreza Kachabi: 0009-0005-2627-942X

Naomi C. Chesler: 0000-0002-7612-5796

* Corresponding author

Email: nchesler@uci.edu (NCC)

**Abstracts**

Computational cardiovascular flow models are highly sensitive to prescribed inlet velocity profiles, yet current approaches often rely either on direct incorporation of high-dimensional imaging-derived velocity fields or simplified analytical velocity-profile formulations. While imaging-based approaches can provide physiologically realistic velocity information, they may also introduce increased preprocessing complexity, imaging noise, and computational burden. In contrast, simplified analytical formulations are computationally efficient but may not fully capture subject-specific flow characteristics. In this study, we propose an uncertainty-aware framework that combines two-dimensional phase-contrast magnetic resonance imaging (2D PC-MRI) with mechanistic velocity-profile formulations to generate subject-specific reduced-order pulmonary artery velocity representations.

Imaging-derived radial velocity distributions were constructed from main pulmonary artery (MPA) PC-MRI data in canine and swine subjects using elliptical radial binning and normalization to enforce physiologically motivated centerline and wall behavior. Power-law and Womersley velocity-profile formulations were fitted to the radial velocity distributions within a Bayesian inference framework while accounting for uncertainty associated with both imaging measurements and model representation. The two formulations were quantitatively compared using regional and global weighted root mean square error (wRMSE) metrics.

Both models demonstrated close agreement with the imaging-derived velocity profiles across subjects. The Womersley formulation showed slightly improved fitting behavior near the vessel wall, whereas the Power-law model demonstrated comparable overall performance despite its simpler structure. No statistically significant differences were observed between the two formulations. Overall, the proposed framework provides low-dimensional, physiologically interpretable, and uncertainty-aware velocity-profile representations that may serve as computationally efficient alternatives for subject-specific cardiovascular flow modeling applications.

## 1. Introduction

Computational cardiovascular models have increasingly been used as informative tools for simulating hemodynamic quantities that may otherwise be difficult or invasive to measure directly in individual patients (Colebank, et al., 2024.; Lechuga et al., 2026). Numerous studies have investigated blood flow dynamics in both the systemic and pulmonary circulations using a variety of approaches, including three-dimensional (3D) computational fluid dynamics (CFD) simulations (Pillalamarri et al., 2021; Zhang et al., 2014), one-dimensional (1D) reduced-order models (Colebank et al., 2021; Duanmu et al., 2019), and coupled multidimensional lumped-parameter frameworks (Zhang et al., 2014). As patient-specific modeling becomes increasingly important for understanding individualized cardiovascular physiology and disease progression, many studies have incorporated patient-derived data into these computational frameworks to improve physiological insight and predictive accuracy (Kheyfets et al., 2015; Steinman, 2002; Zambrano et al., 2021).

Inlet boundary conditions, particularly the assumed velocity-profile distribution, play an important role in cardiovascular flow simulations. Previous studies have demonstrated that downstream hemodynamic quantities can be highly sensitive to the assumed inlet velocity profile, with changes in velocity-profile shape directly influencing wall shear stress (WSS), pressure gradients, recirculation regions, and secondary flow structures (Campbell et al., 2012; Youssefi et al., 2018). Several approaches have been proposed to derive these inflow conditions. Some studies use patient-specific velocity fields obtained from four-dimensional (4D) flow MRI or two-dimensional phase-contrast magnetic resonance imaging (2D PC-MRI) (Bollache et al., 2016; Markl et al., 2012; Stalder et al., 2008). Although these approaches can provide physiologically realistic velocity information, they may also introduce challenges related to image noise, limited spatial resolution, partial-volume effects, increased preprocessing complexity, and can be computationally expensive (Lotz et al., 2002; Morris et al., 2015).

Alternatively, many cardiovascular modeling studies employ simplified analytical velocity-profile formulations, such as Power-law or Womersley-based profiles, because they provide smooth, computationally efficient, and physiologically interpretable inflow representations (Colebank and Chesler., 2024.; Kozitza et al., 2024; McCarthy et al., 2025; Xiong et al., 2011). The Power-law formulation provides a reduced-order representation capable of describing vascular velocity distributions ranging from parabolic-like to blunt flow profiles through variation of the Power-law exponent, while Womersley formulation incorporates pulsatile flow physics derived from analytical solutions of oscillatory viscous flow. However, these simplified assumptions may not fully capture subject-specific flow behavior and are often prescribed uniformly across individuals. Consequently, existing approaches generally rely either on fully prescribed analytical assumptions or direct use of MRI-derived velocity fields, potentially limiting the balance between physiological realism, computational tractability, and robustness to measurement uncertainty.

In the present study, we combine imaging-derived and mechanistic modeling approaches by integrating subject-specific velocity information obtained from 2D PC-MRI at the main pulmonary artery (MPA) with Power-law and Womersley velocity-profile formulations. The pulmonary circulation is a highly pulsatile and compliant vascular system in which accurate characterization of inlet velocity-profile structure may play an important role in subject-specific hemodynamic analysis and cardiovascular flow prediction. Using a Bayesian inference framework, both velocity profiles parameters were estimated directly from imaging-derived radial velocity profiles while accounting for uncertainty associated with observational noise and model representation. Quantifying uncertainty is particularly important because MRI-derived velocity measurements may contain variability related to phase noise, limited spatial resolution, and preprocessing assumptions.

The overall goal of this work is to develop a low-dimensional, physiologically constrained, and uncertainty-aware framework for representing pulmonary artery velocity profiles. Such reduced-order velocity representations may provide improved subject-specific inflow conditions for future cardiovascular CFD simulations, patient-specific modeling, and digital-twin applications while remaining computationally efficient and physiologically interpretable. To our knowledge, this is the first study to integrate imaging-derived pulmonary artery velocity information with uncertainty-aware mechanistic velocity-profile formulations, providing a framework that may be further extended to systemic cardiovascular flow applications. The proposed uncertainty-aware framework may enable future propagation of inlet-condition uncertainty into downstream CFD-based hemodynamic predictions.

## 2. Methods

### 2.1 Image acquisition and ROI extraction

Data included 2D PC-MRI of the MPA, consisting of both magnitude and phase, from five healthy male canines (12.3 ± 0.9 kg body weight) (Bellofiore et al., 2013) and four female sham swine (30 ± 20.6 kg body weight) (Altieri et al., 2025). We denote canine subjects 1–5 as C1–C5 and swine subjects 1–4 as S1–S4, respectively. Only systolic images were used, as these provided the most contrast. Using ImageJ, the MPA was identified from the magnitude image, and its borders were approximately delineated. The coordinates and corresponding image intensities within this approximate region of interest (ROI) were then extracted and stored. After ROI extraction, all subsequent analyses including ellipse fitting, radial binning, normalization, Bayesian inference and uncertainty quantification were performed in Python and are available at https://github.com/AmirrezaKachabi .

Since the vessel cross-section was not perfectly circular in the image plane, an ellipse was fit to the manually selected ROI pixels as shown in Fig. 1. The ellipse was parameterized by its center $(x_0\,, y_0)$, semi-major and semi-minor axes $(a\,, b)$, and rotation angle $\psi$. For each ROI pixel, a normalized elliptical radial coordinate was computed as

$$r = \sqrt{\left(\frac{x'}{a}\right)^2 + \left(\frac{y'}{b}\right)^2} \tag{1}$$

where $x'$ and $y'$ are the coordinates after translating the pixel by the ellipse center and rotating by $\psi$. Pixels with $r \leq 1$ were retained for analysis, and $r$ was treated as the normalized elliptical radius, where $r = 0$ corresponds to the vessel center and $r = 1$ corresponds to the vessel wall. Finally, the same ROI pixels were sampled from the phase image, and phase correction was applied to ensure a positive centerline velocity signal. This step accounted for possible sign inversion in the phase-contrast velocity encoding caused by imaging orientation and flow direction (Lotz et al., 2002).

**Figure 1**

## 2.2 Radial binning and normalization

The corrected phase values were grouped into radial bins defined over the normalized elliptical radius $0 \leq r \leq 1$. Multiple bin numbers were evaluated to assess the effect of bin resolution on velocity profiles smoothness and model fitting stability. For each radial bin, the mean corrected phase value, standard deviation, and number of pixels were computed. Based on physiological and fluid dynamic assumptions, velocity near the vessel wall is expected to approach zero due to the no-slip condition, while the centerline region is expected to show the highest velocity at peak systole. Because the image-derived phase data contained measurement noise and spatial variability, the binned radial profile was normalized using a center-wall normalization:

$$I_{\text{norm}}(r) = \frac{I(r) - I_{\text{edge}}}{I_{\text{center}} - I_{\text{edge}}} \tag{2}$$

where $I_{\text{center}}$ was the median binned value for $r < 0.2$ and $I_{\text{edge}}$ was the median binned value for $r > 0.9$.

The normalized phase/intensity values were constrained to the range [0,1]. This normalization imposed a center-to-wall interpretation in which the vessel center represented the region of highest relative velocity while the wall approached zero velocity. Since the number of radial bins could impact the final estimated velocity profile parameters, a sensitivity analysis was performed across all subjects and velocity profile models using radial bin counts of 5, 10, 12, 15, 18, 20, 25, 30, 40, and 50. Based on profile smoothness and model fitting performance, 15 radial bins were selected and fixed across all nine subjects for the final analysis as a balance between spatial resolution and noise stability.

## 2.3 Velocity profile models

### 2.3.1 Power-law velocity profile

The normalized radial profile was first modeled using a Power-law velocity profile:

$$U(r) = \bar{U}\frac{\gamma+2}{\gamma}(1-r^{\gamma}) \tag{3}$$

where $r$ is the normalized radial coordinate ranging from 0 at the vessel center to 1 at the vessel wall, $\gamma$ controls profile bluntness, and $\bar{U}$ represents the mean velocity scaling parameter. A preliminary estimate of $\bar{U}$ was computed directly from the normalized profile using area integration:

$$Q \approx 2\pi ab \int_0^1 r I_{norm}(r)\, dr \tag{4}$$

$$\bar{U}_{\text{data}} = \frac{Q}{\pi ab} = 2\int_0^1 r I_{norm}(r)\, dr \tag{5}$$

This value was later used as the center of the prior distribution for Bayesian inference.

2.3.2 Womersley velocity profile

The radial profile was also modeled using a Womersley-based pulsatile flow formulation:

$$U(r) = A\, Re\left\{\left[\left(1 - \frac{J_0(\lambda r)}{J_0(\lambda)}\right) e^{i\phi}\right]\right\} + C \tag{6}$$

where $Re$ denotes the real part of the complex-valued Womersley solution, r is the normalized radial coordinate, $A$ is the amplitude, $\phi$ is the phase parameter, $J_0$ is the zeroth-order Bessel function and $\lambda = i^{3/2}\alpha$ where $\alpha$ is the Womersley number. The offset parameter C was fixed to zero because the profile had already been baseline-corrected and normalized. The physiological Womersley number was estimated from subject-specific vessel geometry and cardiac cycle information:

$$\alpha_{\text{phys}} = R_{\text{phys}}\sqrt{\frac{\omega}{\nu}} \tag{7}$$

where $R_{phys}$ is the physical vessel radius, defined as the area-equivalent radius of the fitted ellipse $\left(\sqrt{ab}\right)$, $\omega = 2\pi/T$ is the angular frequency derived from the cardiac cycle duration $T$, and $\nu$ is the kinematic viscosity. The resulting subject-specific estimate, $\alpha_{\text{phys}}$, was subsequently used as the center of the prior distribution during Bayesian inference of the Womersley velocity profile.

2.4 Data uncertainty estimation

We separated the data uncertainty coming from the radial binning and normalization processes and ultimately defined a single term, $\sigma_{\text{data}}$, representing the total observational uncertainty prior to Bayesian parameter inference.

For each radial bin, the standard error of the mean (SEM) was computed as:

$$\sigma_{\text{SEM},i} = \frac{\sigma_i}{\sqrt{N_i}} \tag{8}$$

where $\sigma_i$ is the standard deviation within bin $i$, and $N_i$ is the number of pixels in that bin. Because the data were normalized to satisfy physiological boundary conditions at the vessel wall and center, uncertainty associated with these assumptions needed to be accounted for. Since normalization depends on estimated center and wall reference values, uncertainty in these assumptions propagates throughout the normalized profile. To account for this effect, bootstrap resampling was performed. Within each radial bin, pixels were repeatedly resampled with replacement, and the normalization procedure was recomputed for each realization. The variability across the bootstrapped normalized profiles was used to estimate normalization-induced uncertainty:

$$\sigma_{\text{norm},i} = \text{std of bootstrapped normalized values}$$

This procedure accounts for uncertainty in the assumptions that the wall region represents minimal velocity and that the vessel center represents maximal velocity. Finally, the total observational uncertainty for each radial bin was computed as:

$$\sigma_{\text{data},i} = \sqrt{\sigma_{\text{SEM},i}^2 + \sigma_{\text{norm},i}^2} \tag{9}$$

This produced a heteroscedastic uncertainty profile in which different radial locations were assigned different levels of confidence.

2.5 Parameter estimation

We used a Bayesian framework for parameter inference, where the model parameters were treated as random variables. Prior parameter distributions were combined with the likelihood function to estimate the posterior distribution conditioned on the observed data. This formulation follows Bayes' theorem:

$$p(\theta \mid I) = \frac{L(I \mid \theta)\, p(\theta)}{p(I)} \tag{10}$$

where $I$represents the measured normalized radial velocity profile, $L(I \mid \theta)$is the likelihood function, $p(\theta)$ is the prior distribution of the parameters, and $p(I)$ is the evidence or normalization factor. The posterior distribution was approximated using Markov chain Monte Carlo (MCMC) sampling. For the Power-law velocity model, the inferred parameter set was defined as $\theta_P = \{\gamma, \bar{U}\}$ and for the Womersley velocity profile, the inferred parameter set was defined as $\theta_W = \{\alpha, A, \phi\}$. The parameter ranges used for posterior sampling are summarized in Table 1.

**Table 1** Parameters ranges for Bayesian inference.

| Model | Parameter | Description | Range |
|---|---|---|---|
| Power-law | $\gamma$ | Profile bluntness exponent | $[2, 9]$ |
| Power-law | $\bar{U}$ | Mean normalized velocity | $[0.5\,\bar{U}_{data},\ 1.5\,\bar{U}_{data}]$ |
| Womersley | $\alpha$ | Womersley number | $[\alpha_{phys} - 8,\ \alpha_{phys} + 8]$ |
| Womersley | $A$ | Velocity amplitude scaling | $[0.2\,I_{max},\ 2.5\,I_{max}]$ |
| Womersley | $\phi$ | Phase shift | $[-\pi, \pi]$ |

The normalized radial profile at each radial bin $i$was modeled as

$$I_i \sim \mathcal{N}\left(U(r_i;\theta), \sigma_{\text{obs},i}^2\right) \tag{11}$$

where $U(r_i;\theta)$denotes the model prediction and $\sigma_{\text{obs},i}$represents the total uncertainty at radial location $r_i$. The observational uncertainty was defined as:

$$\sigma_{\text{obs},i}^2 = \sigma_{\text{data},i}^2 + \sigma_{\text{model}}^2 \tag{12}$$

where $\sigma_{\text{data},i}$ includes both the standard error of the normalized radial-bin intensity and the uncertainty introduced during normalization, while $\sigma_{\text{model}}$ represents a global residual model discrepancy term shared across all radial locations. Because the likelihood variance varies across radial bins, radial regions with larger uncertainty naturally contribute less to the posterior parameter estimates.

The physical model parameters were assigned flat uniform priors except for $\bar{U}$ in the Power-law model and $\alpha$ in the Womersley model. For the Power-law model, $\bar{U}$ was assigned a data-driven Gaussian prior centered at the preliminary velocity estimate obtained from the normalized radial profile, $\bar{U} \sim \mathcal{N}(\bar{U}_{\text{data}}, (0.25\,\bar{U}_{\text{data}})^2)$. For the Womersley model, $\alpha$ was assigned a Gaussian prior centered at the subject-specific physiological estimate computed from vessel radius and cardiac cycle information, $\alpha \sim \mathcal{N}\left(\alpha_{\text{phys}}, 2^2\right)$. The residual model discrepancy term, $\sigma_{\text{model}}$, was assigned a weakly informative half-normal prior with scale parameter 0.03, corresponding to approximately 3% of the normalized profile range.

The likelihood function was defined as:

$$L(I \mid \theta) = \frac{1}{(2\pi)^{N/2} \mid \Sigma_{\text{obs}} \mid^{1/2}} \exp\left[-\frac{1}{2}(I - U(\theta))^T \Sigma_{\text{obs}}^{-1} (I - U(\theta))\right] \tag{13}$$

where $N$ is the number of radial bins and the observational covariance matrix was defined as

$$\Sigma_{\text{obs}} = \text{diag}\left(\sigma_{\text{obs},1}^2, \ldots, \sigma_{\text{obs},N}^2\right) \tag{14}$$

Posterior distributions were sampled using a Delayed Rejection Adaptive Metropolis (DRAM) algorithm implemented in Python. The sampler employed Gaussian random-walk proposals with adaptive covariance updates based on the empirical covariance of the sampled chains. When a proposed transition was rejected, a second smaller proposal was attempted to improve local exploration and sampling efficiency. The DRAM sampler was run for 60,000 iterations, with the first 10,000 iterations treated as burn-in and discarded from the posterior analysis. Convergence was assessed using trace plots, posterior density stability, and the Gelman–Rubin $\hat{R}$ diagnostic across four independent chains, with values of $\hat{R} < 1.05$ considered consistent with convergence (Vehtari et al., 2021). Posterior summaries, including posterior means, maximum a posteriori (MAP) estimates, posterior medians, and 95% credible intervals, were computed from the remaining posterior samples. To quantify uncertainty in the fitted velocity profiles, both credible and prediction intervals were computed. Credible intervals (CI) represent the range within which the model parameters or fitted velocity profile are likely to lie with a specified posterior probability conditioned on the observed data. In contrast, prediction intervals (PI) represent the range likely to contain a future observation by additionally accounting for observational uncertainty and residual model discrepancy and are therefore generally wider. Uncertainty intervals for the velocity profiles were constructed using posterior samples retained after burn-in. Specifically, 3,000 posterior samples randomly drawn from the converged MCMC chains were used to generate the reported credible and prediction intervals for both the Power-law and Womersley velocity profiles. Prediction intervals incorporated the uncertainty

$$\sigma_{\text{pred}}(r) = \sqrt{\sigma_{\text{data}}(r)^2 + \sigma_{\text{model}}^2} \tag{15}$$

such that the resulting prediction intervals reflect both posterior parameter uncertainty and the expected variability in normalized velocity observations.

2.6 Model comparison

We performed a comparison between the Power-law and Womersley velocity profiles to determine which model better explained the observed data. Because each radial bin represents a different annular area of the vessel cross-section, weighted error metrics were used for model comparison. The global weighted root mean square error (wRMSE) was defined as:

$$\text{wRMSE} = \sqrt{\frac{\sum_i w_i \left(I_i - \hat{I}_i\right)^2}{\sum_i w_i}} \tag{16}$$

where $w_i = N_i$ corresponds to the number of pixels within each radial bin, $I_i$ represents the normalized observed data, and $\hat{I}_i$ denotes the model prediction. In addition to global wRMSE, regional wRMSE values were evaluated to investigate how the different velocity-profile models represented distinct portions of the normalized radial profile individually. Because each profile was represented using 15 radial bins, model performance was evaluated using bin-index-based regions to ensure that each region contained a sufficient and consistent number of binned observations across subjects. The final three bins closest to the vessel boundary were defined as the near-wall region, the preceding three bins were defined as the transitional region, and the remaining inner bins were defined as the center region. Figure. 2 shows separated radial bins for a representative subject based on three different regions.

**Figure 2**

2.6 Statistical analysis

Statistical analyses were performed to compare model-derived and data-derived parameter estimates, as well as regional wRMSE values between the Power-law and Womersley models. Paired comparisons between model-derived and data-derived parameters were performed using paired t-tests after verification of normality in paired differences. Comparisons of wRMSE values between the Power-law and Womersley models were performed using the Wilcoxon signed-rank test because both models were evaluated on the same subjects, resulting in paired observations. Statistical significance was defined as $p < 0.05$.

## 3. Results

### 3.1 Velocity Profile Fits and Uncertainty Quantification

Power-law model fitting for canine and swine subjects are shown in Figures 3 and 4, respectively, while Womersley model fits are shown in Figures 5 and 6, respectively. We also present 95% credible and prediction intervals derived from 3,000 random posterior samples for all subjects. As shown in Figures 3–6, both models were able to closely capture the measured normalized velocity profiles. Credible and prediction interval widths varied across subjects and models. In most subjects, uncertainty generally increased from the vessel center toward the wall, particularly in regions with steep velocity gradients and reduced pixel support. However, C5, S1, and S4 demonstrated broader uncertainty across larger portions of the vessel lumen, particularly within the swine datasets and Womersley fits, reflecting increased observational variability and model flexibility. Due to its Bessel-function-based formulation, the Womersley model showed improved near-wall agreement compared to the Power-law model, which tended to overestimate velocity in that region. However, this increased flexibility was also associated with broader credible intervals and greater fitting variability, particularly within the transitional and near-wall regions. Despite the increased noise, both modeling approaches remained capable of recovering physiologically consistent blunt velocity distributions across all subjects. We also present supplementary figures (Figures Sup_1 and Sup_2) showing model predictions evaluated at the posterior mode parameters for each subject to better visualize the differences between the Power-law and Womersley fits.

**Figures 3-6**

### 3.2 Regional wRMSE Comparison Between Velocity Profiles

The wRMSE box-and-whisker plots for all subjects and vessel regions are shown in Figure. 7 Globally, the Womersley model demonstrated slightly lower wRMSE compared to the Power-law model $(8.39 \pm 4.08) \times 10^{-2}$ vs. $(9.10 \pm 1.89) \times 10^{-2}$. However, regionally, performance varied between models. The Power-law model generally produced lower errors within the central lumen and transitional regions, whereas the Womersley model demonstrated improved agreement near the vessel wall. None of these regional or global differences reached statistical significance. Variability in fitting error was consistently larger for the Womersley model, particularly in the global and near-wall regions, indicating greater inter-subject heterogeneity. This behavior was consistent with the broader credible intervals observed in the Womersley fits shown in Figures. 5 and 6.

Species-specific trends were also observed. In the global region, canine and swine subjects demonstrated relatively similar error distributions for the Power-law model, whereas the Womersley model showed greater variability and generally higher errors among swine subjects. Within the center region, both species again demonstrated relatively similar Power-law fitting behavior, except for S3, which exhibited elevated center-region error. In contrast, the Womersley center-region fits showed substantial overlap between canine and swine subjects without a clear species-dependent trend. Within the transitional region, both species demonstrated comparable fitting behavior across the two models. However, in the near-wall region, canine subjects generally clustered at lower Womersley error values compared to swine subjects, suggesting more consistent near-wall fitting behavior within the canine datasets.

**Figure 7**

### 3.3. Marginal Posterior Densities and Model Parameters

Power-law marginal posterior distributions are shown in Figures. 8 and 9, respectively, while Womersley marginal posterior distributions are shown in Figures. 10 and 11, respectively. In all figures, the y-axis represents the probability density function (PDF), indicating the probability density associated with each parameter value, while the x-axis shows the range of the corresponding parameter. Note that each PDF integrates to unity. For visualization purposes, posterior probability densities were estimated using kernel density estimation (KDE) applied to the retained MCMC samples. Overall, the marginal posterior distributions were predominantly unimodal across all subjects and parameters, indicating well-constrained parameter estimates with no evidence of multimodal behavior. Most parameters demonstrated relatively compact posterior distributions with only mild skewness. Among all parameters, the phase parameter $\varphi$ consistently exhibited a broader left-skewed tail across subjects, suggesting greater asymmetry and uncertainty toward lower phase values. Table. 2 contains the model parameters estimated at the MAP for each subject, along with the data-derived mean velocity values and physiological Womersley numbers. Overall, the inferred $\gamma$ values corresponded to moderately blunt velocity distributions, while the inferred $\alpha$ values remained within the relatively high-$\alpha$ regime across subjects.

**Figures 8-11**

**Table 2.** Subject-specific posterior MAP parameter estimates for the Power-law and Womersley models, along with data-derived mean velocity values and physiological Womersley numbers. Mean ± standard deviation values across all subjects are also provided. For the Power-law model, the estimated MAP mean velocity values were significantly higher than the data-derived mean velocity values ($p < 0.005$). For the Womersley model, the physiology-derived Womersley numbers were significantly higher than the estimated MAP Womersley numbers ($p < 0.005$).

| ID | $\gamma$ | $\bar{U}_{MAP}$ | $\bar{U}_{data}$ | $\alpha_{MAP}$ | $\alpha_{phys}$ | $A_{MAP}$ | $\phi_{MAP}$ |
|---|---|---|---|---|---|---|---|
| | | *** | *** | *** | *** | | |
| C1 | 6.25 | 0.76 | 0.72 | 15.4 | 19.3 | 2.01 | 1.05 |
| C2 | 6.9 | 0.78 | 0.74 | 13.2 | 16.7 | 1.67 | 0.94 |
| C3 | 5.28 | 0.73 | 0.71 | 13.9 | 14.7 | 1.61 | 0.91 |
| C4 | 5.36 | 0.72 | 0.71 | 12.3 | 14.3 | 1.58 | 0.90 |
| C5 | 5.77 | 0.74 | 0.74 | 14.2 | 15.2 | 1.66 | 0.94 |
| S1 | 5.59 | 0.74 | 0.68 | 15.3 | 18.0 | 1.99 | 1.06 |
| S2 | 7.33 | 0.79 | 0.76 | 19.1 | 21.5 | 2.03 | 1.06 |
| S3 | 5.75 | 0.74 | 0.70 | 20.6 | 22.2 | 2.03 | 1.07 |
| S4 | 6.08 | 0.75 | 0.73 | 18.7 | 24.4 | 2.07 | 1.08 |
| Mean ± SD | $6.03 \pm 0.65$ | $0.75 \pm 0.02$ | $0.72 \pm 0.02$ | $15.9 \pm 2.8$ | $18.5 \pm 3.4$ | $1.85 \pm 0.20$ | $1.00 \pm 0.07$ |

## 4. Discussion

In this study, we characterized 2D PC-MRI-derived velocity profiles in the MPA using both Power-law and Womersley velocity-profile formulations in canine and swine subjects. A Bayesian inference framework was employed to estimate subject-specific model parameters while quantifying uncertainty associated with observational noise, imaging variability, and model representation. To assess the ability of these mechanistic formulations to reproduce MRI-derived pulmonary artery velocity distributions, the two velocity-profile models were quantitatively compared across all subjects and vessel regions. Overall, the proposed framework enabled the construction of subject-specific reduced-order representations of pulmonary artery velocity profiles while accounting for physiological constraints and uncertainty from both imaging measurements and model assumptions.

### 4.1 Bayesian Parameter Estimation

Parameter estimation involves assessing the consistency between subject-specific data and the predictions generated by the proposed models. Previous studies in cardiovascular modeling have estimated model parameters using either frequentist approaches (Kim et al., 2026; Zambrano et al., 2018) or Bayesian inference frameworks (Kachabi et al., 2025; Paun et al., 2020). Although both methodologies provide mechanisms for uncertainty quantification, Bayesian approaches yield full posterior parameter distributions, enabling a more comprehensive characterization and interpretation of uncertainty compared with the point estimates and confidence intervals typically obtained from frequentist methods (Smith, 2024). In addition, Bayesian frameworks naturally allow incorporation of prior physiological or experimental knowledge into the inference process, which can be particularly useful when working with noisy imaging-derived measurements or limited datasets.

As shown in Figures 3–6, both velocity-profile formulations demonstrated close agreement with the imaging-derived radial velocity data across nearly all subjects. Overall, increased uncertainty intervals were consistently observed near the vessel wall regions. This behavior is consistent with a previous PC-MRI study, which has reported that limited spatial resolution and partial-volume effects can introduce increased uncertainty and inaccuracies in velocity estimation near vessel boundaries (Cibis et al., 2016). Swine subjects, particularly S1 and S4, demonstrated broader prediction intervals compared with the remaining subjects, which may reflect increased noise and variability within their imaging datasets. Figures 8–11 further illustrate the posterior parameter distributions for all subjects and model parameters. Overall, the posterior densities remained predominantly unimodal without substantial long-tailed behavior or multimodal structure, suggesting stable parameter inference and generally well-constrained model fits across subjects.

4.2 Power-law vs. Womersley

Several studies have investigated the influence of different velocity-profile formulations in cardiovascular flow modeling. San and Staples (San & Staples, 2012) compared Womersley-based profiles with classical parabolic and blunt velocity assumptions (Power-law profile with $\gamma = 9$) in reduced-order blood flow models and concluded that the Womersley formulation improved physiological realism in pulsatile flow simulations. Ponzini et al. (Ponzini et al., 2010) used 2D PC- MRI data acquired from ten healthy young volunteers across three arterial regions, including the abdominal aorta, common carotid artery, and brachial artery. Their study compared Womersley-based formulations with simplified parabolic and flat velocity assumptions and demonstrated that incorporating Womersley flow physics improved estimation of pulsatile blood-flow characteristics, while simplified velocity assumptions were insufficient to fully represent the measured vascular flow profiles. In the present study, we used 2D PC-MRI data acquired in the MPA to compare Power-law and Womersley velocity-profile formulations across five canine and four swine subjects and across different vessel regions. As shown in Figure. 7, the Womersley model demonstrated slightly lower global weighted RMSE compared with the Power-law model. This behavior was likely influenced by the increased contribution of the near-wall regions within the weighted error calculation. Regional analysis further demonstrated that the two models performed similarly within the central lumen region, whereas the Power-law formulation showed slightly improved agreement within the transitional region. In contrast, the Womersley formulation produced lower wRMSE values near the vessel wall, suggesting improved representation of near-wall pulsatile flow behavior. However, none of these regional or global differences reached statistical significance.

4.2 Velocity Profile Parameters

Previous studies have employed different velocity-profile assumptions within cardiovascular flow modeling frameworks. In the context of pulmonary circulation, several studies have assumed blunt or flat velocity distributions for computational simulations (Kozitza et al., 2024.; Smith et al., 2001). In our previous works (Kachabi et al., 2023, 2024), we used the same 2D PC-MRI canine dataset but employed a simplified approach in which the MPA was approximated using a circular geometry and a fixed Power-law exponent of $\gamma = 5$ was prescribed across all subjects. The selection of precise $\gamma$ is important because the assumed velocity-profile shape can substantially influence downstream hemodynamic predictions, including WSS, which has been associated with endothelial dysfunction and vascular remodeling (Allen et al., 2023). In cardiovascular physiology, abnormal WSS environments have been linked to pathological vascular conditions and altered vascular function (Li et al., 2009), emphasizing the importance of accurately characterizing inlet velocity distributions for reliable hemodynamic simulations and potential clinical interpretation.

In the present study, we extended this framework by fitting an elliptical representation of the MPA, which provided a more physiologically realistic approximation than the previously assumed circular geometry, and directly estimating subject-specific velocity-profile parameters from imaging-derived radial velocity data using Bayesian inference to quantify uncertainty in the inferred model parameters. The inferred Power-law exponents were generally centered around $\gamma \approx 6$ as reported in Table 2, which remained relatively close to the simplified assumptions used in our previous study while differing substantially from flatter blunt-profile assumptions such as $\gamma = 9$. These findings suggest that although simplified blunt velocity assumptions may provide computational convenience, they may not fully represent the subject-specific pulmonary artery velocity distributions observed in the imaging data. The inferred mean velocity parameter $\bar{U}$ remained relatively consistent across subjects for both velocity-profile formulations. Compared with the imaging-derived prior estimates, the MAP values of $\bar{U}$ were significantly higher, suggesting that the optimized mechanistic profiles favored slightly elevated mean velocity magnitudes to better reproduce the measured radial velocity distributions.

Similar studies have demonstrated that assumed velocity-profile formulations can substantially influence estimated pulsatile flow characteristics and derived hemodynamic quantities (Ponzini et al., 2010; Youssefi et al., 2018). In addition, limitations associated with PC-MRI spatial resolution and velocity smoothing may contribute to differences between imaging-derived estimates and optimized mechanistic representations (Bochert et al., 2025). The inferred Womersley numbers demonstrated relatively consistent behavior across subjects, with an average MAP estimate of approximately $\alpha \approx 15.9$. Such moderately high Womersley numbers are physiologically consistent with large-vessel pulsatile hemodynamics in the MPA, where unsteady inertial effects dominate over viscous diffusion during the cardiac cycle (Nichols and McDonald, 2011). Under these conditions, velocity profiles tend to exhibit flatter core flow regions together with stronger near-wall pulsatile gradients and localized phase-dependent curvature. This behavior was qualitatively consistent with the fitted Womersley profiles observed in the Figures 5 and 6, particularly the increased near-wall curvature and oscillatory behavior relative to the smoother monotonic Power-law representations. The relatively blunt velocity distributions observed in both formulations may therefore reflect the underlying high-Womersley pulsatile flow structure of the pulmonary circulation. Also, the MAP estimates of the Womersley number $\alpha$ were significantly lower than the physiology-derived prior estimates used during Bayesian inference. This behavior may suggest that the optimized velocity profiles favored smoother or less oscillatory radial velocity structures compared with the idealized pulsatile assumptions represented by the prior formulation. Differences between prior and inferred $\alpha$ values may additionally reflect limitations associated with imaging resolution, profile normalization, and the simplified cylindrical assumptions inherent to the classical Womersley model. As reported in Table 2, the amplitude parameter $A$ demonstrated mild inter-subject variability, suggesting relatively consistent pulsatile profile magnitudes across the analyzed pulmonary artery datasets. In contrast, the phase parameter $\phi$ remained highly consistent across subjects, with MAP estimates centered near $\phi \approx 1\ (rad)$. This relatively small variability may reflect similar pulsatile timing characteristics and image acquisition conditions among subjects.

## 5. Limitations

Our study has several limitations. First, the sample size for both canine and swine subjects was relatively small. Inclusion of a larger cohort would strengthen the statistical analysis and improve the generalizability of the findings. Second, several datasets, particularly within the swine subjects, contained increased imaging noise and variability, which may have influenced the inferred fitting parameters and associated uncertainty estimates. Third, the radial velocity profiles were normalized to enforce physiologically motivated centerline and near-wall boundary behavior. Although uncertainty associated with the normalization procedure was partially incorporated through the $\sigma_{\text{norm}}$ term, the full impact of the normalization process on the final inferred parameter distributions may not be completely captured. Fourth, a fixed radial bin count of 15 was used across all subjects to maintain methodological consistency and enable regional comparisons. However, individual subjects may potentially have been better represented using slightly different radial bin resolutions depending on image quality and spatial flow complexity, which could have modestly influenced the resulting parameter estimates. Lastly, the classical Womersley formulation used in this study assumes simplified axisymmetric pulsatile flow conditions, which may not fully capture complex secondary flow structures, vessel curvature, and flow asymmetry within the main pulmonary artery.

## 6. Conclusions

In this study, we proposed an uncertainty-aware framework for cardiovascular velocity-profile modeling that bridges the gap between fully imaging-derived velocity fields and simplified prescribed analytical velocity-profile formulations. Rather than directly utilizing high-dimensional velocity fields obtained from advanced imaging techniques such as 4D flow MRI or assuming generic velocity-profile formulations, the proposed framework integrates 2D PC-MRI-derived velocity information with mechanistic Power-law and Womersley velocity-profile models. Bayesian inference was used to estimate subject-specific model parameters while quantifying uncertainty associated with both imaging-derived measurements and model representation, enabling potential propagation of uncertainty into downstream cardiovascular flow simulations. Comparison of the fitted velocity-profile formulations demonstrated that both the Power-law and Womersley models provided comparable agreement with the imaging-derived pulmonary artery velocity profiles across subjects. Although the Womersley formulation incorporates more complex pulsatile flow physics and demonstrated improved near-wall fitting behavior, it did not demonstrate a statistically significant improvement over the simpler Power-law formulation. These findings suggest that relatively simple reduced-order velocity-profile representations may provide physiologically realistic and computationally efficient alternatives for subject-specific cardiovascular flow modeling applications.

## Acknowledgements

This work was supported by the National Institutes of Health NIBIB grants R01HL147590 and by the Edwards Lifesciences Foundation Graduate Fellowship. The authors would like to thank Sofia Altieri Correa for her valuable discussions and contributions to the development of the study methodology. The content is solely the responsibility of the authors and does not necessarily represent the official views of the NIH.

## Ethics declarations

This study did not involve new animal experiments. All procedures were approved by the University of Wisconsin-Madison Institutional Animal Care and Use Committee.

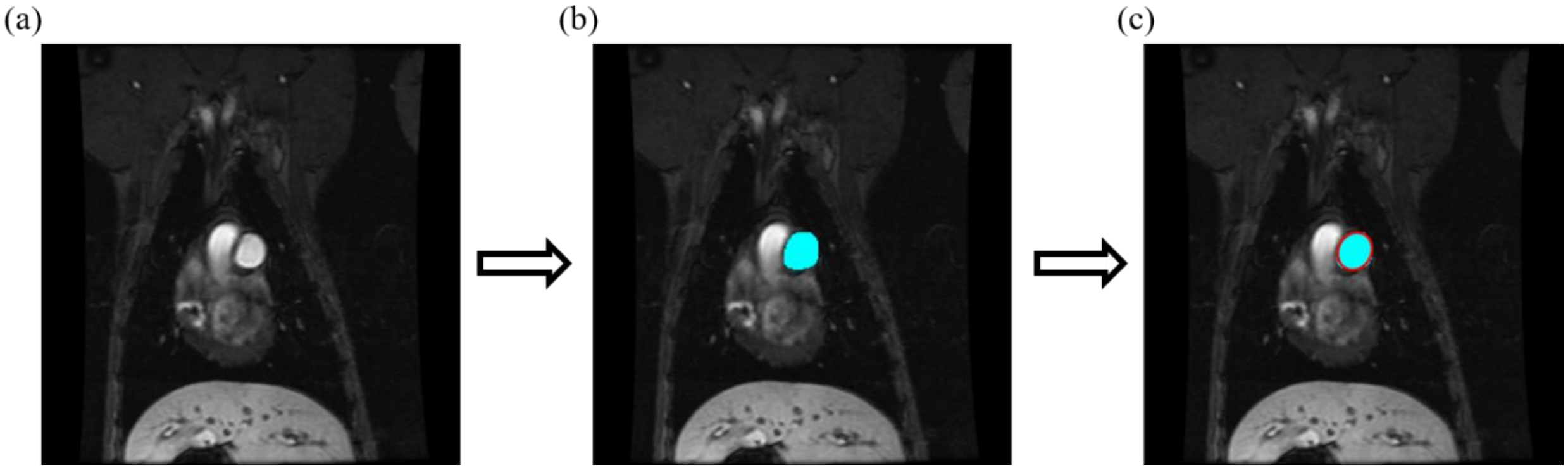

**Figure 1.** Workflow for MPA identification and ellipse fitting from 2D PC-MRI magnitude images. (a) Representative systolic magnitude image showing the visible MPA cross-section. (b) Approximate localization and manual ROI selection of the MPA. (c) Ellipse fitted to the selected ROI to approximate the vessel cross-sectional geometry for subsequent radial profile analysis.

https://doi.org/10.1002/cnm.2625

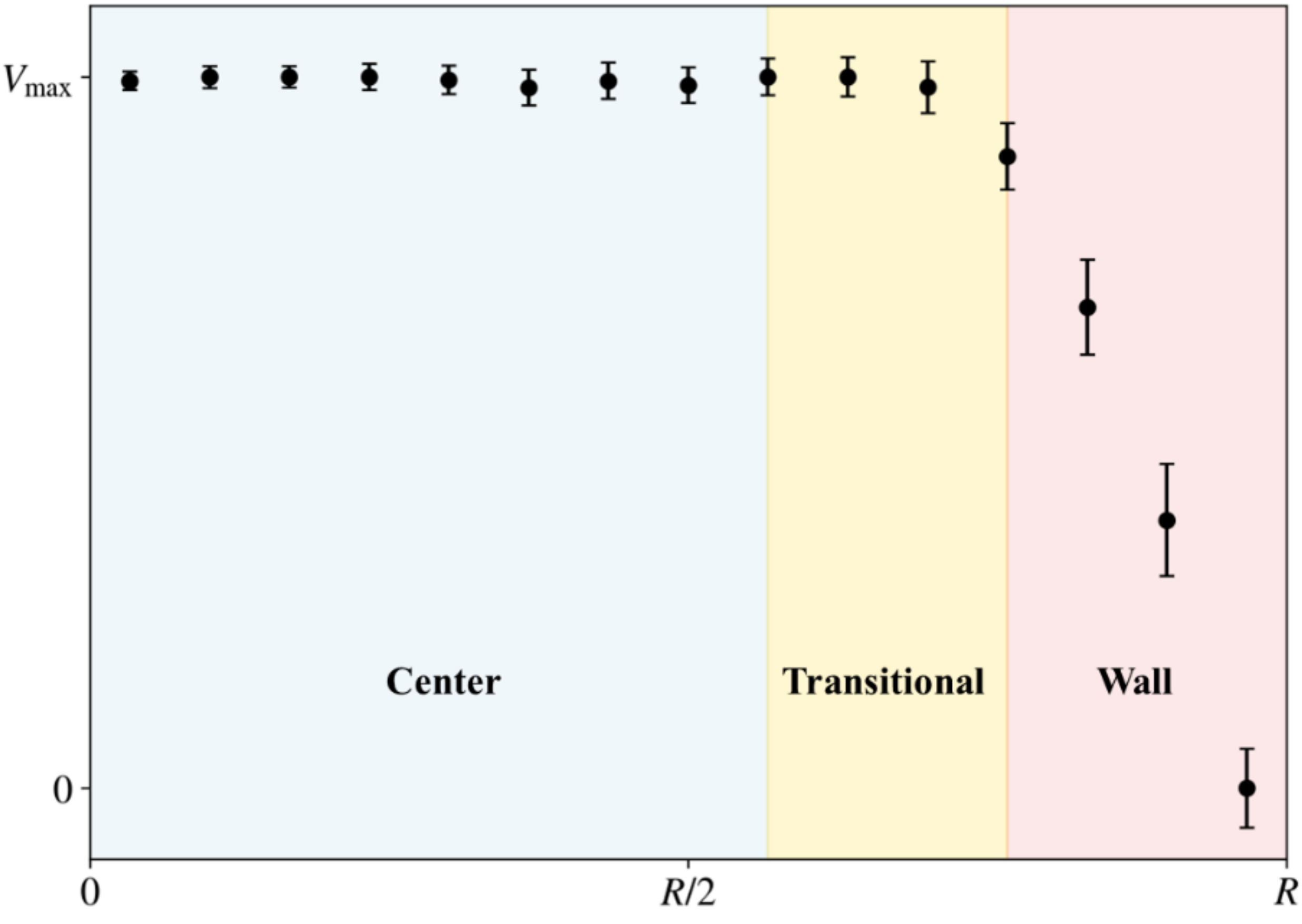

**Figure 2.** Representative subject showing the separation of the normalized radial profile into center (light blue), transitional (yellow), and near-wall (pink) regions based on radial-bin location.

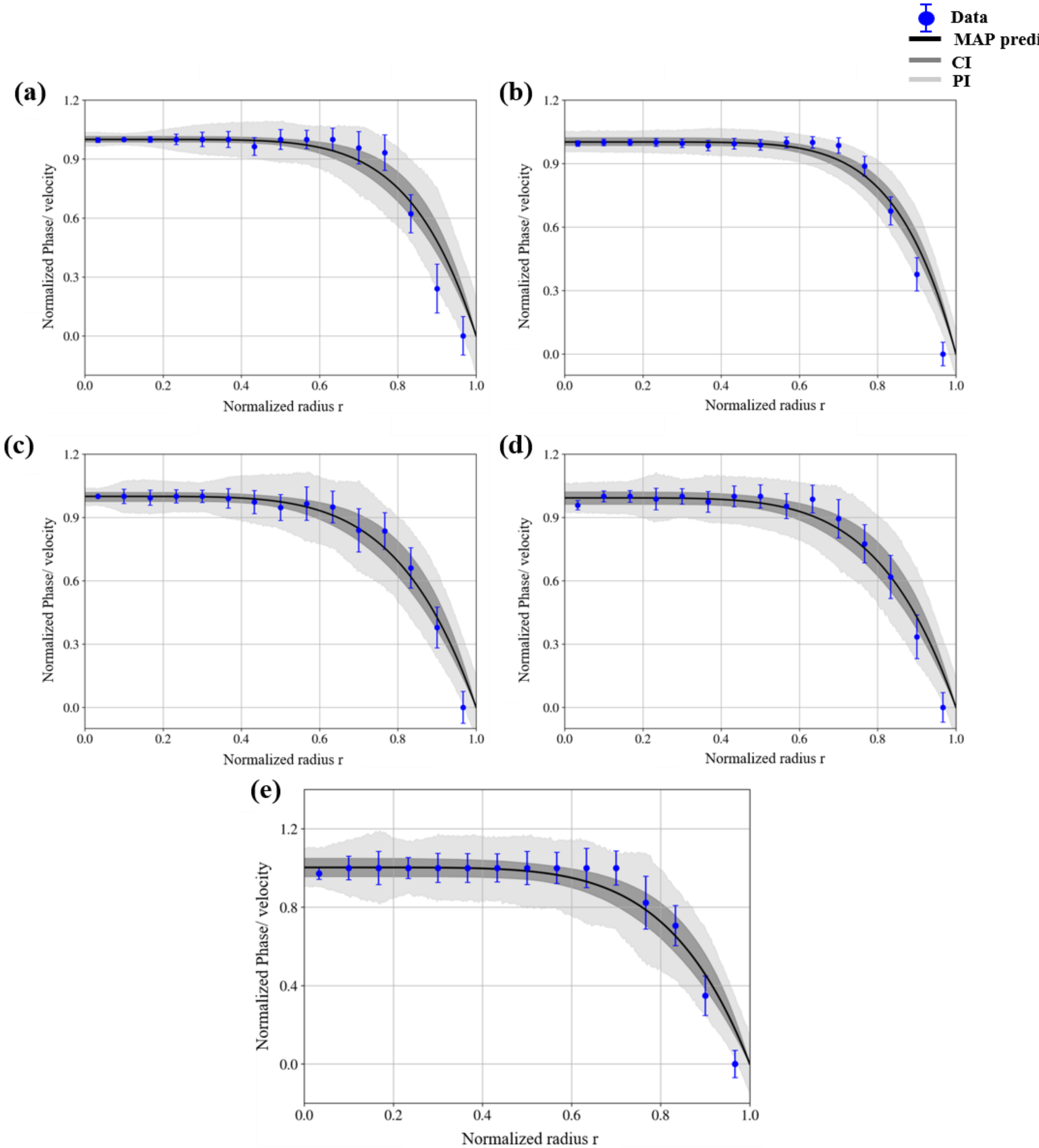

**Figure 3.** Power-law fitting results for all canine subjects. The figure shows the binned radial velocity data with error bars (blue), along with the posterior model prediction evaluated at the MAP estimate (black solid line), the 95% credible interval (dark gray), and the 95% prediction interval (light gray). Subplots are organized as follows: (a) C1, (b) C2, (c) C3, (d) C4, and (e) C5.

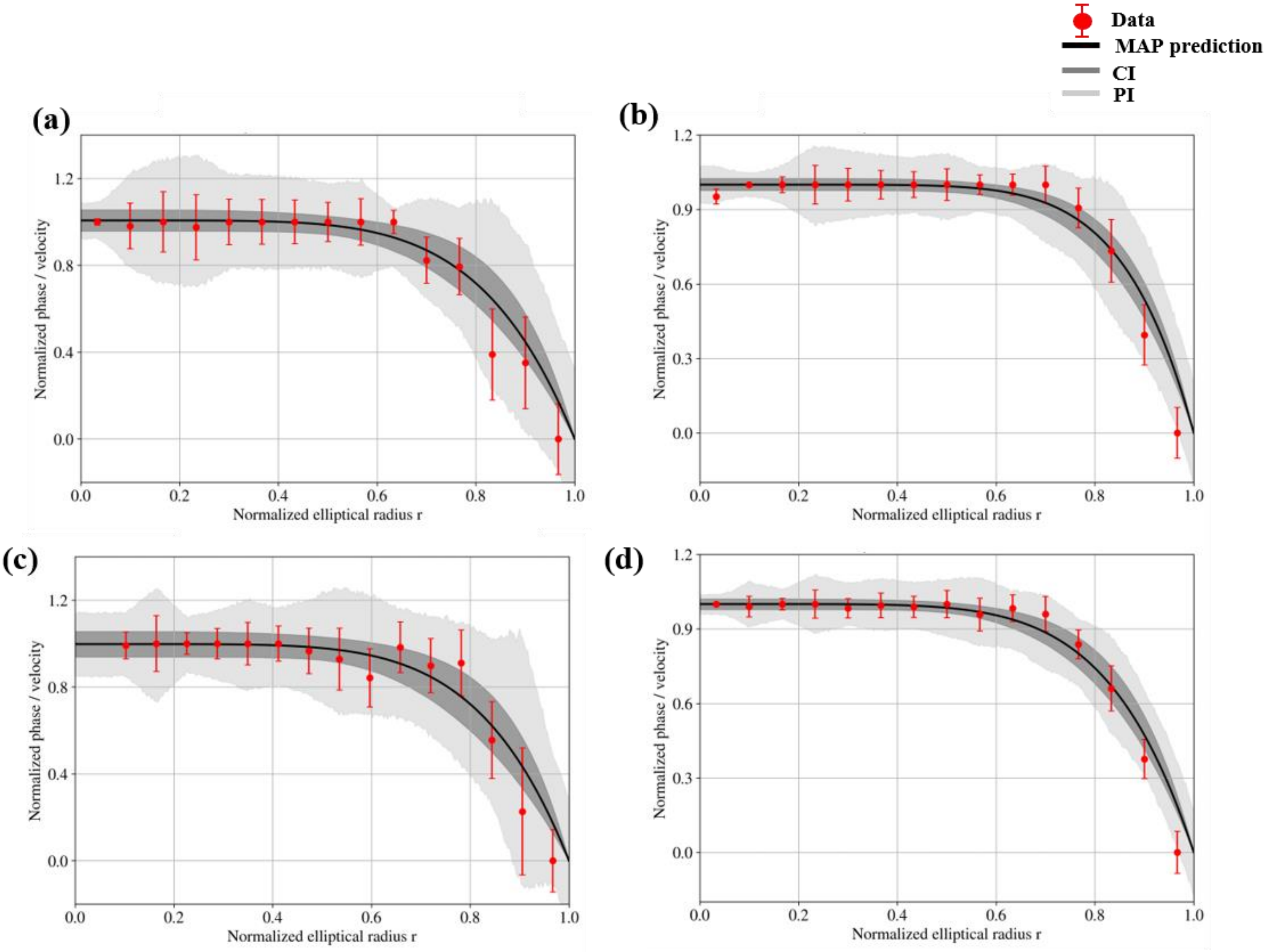

**Figure 4.** Power-law fitting results for all swine subjects. The figure shows the binned radial velocity data with error bars (red), along with the posterior model prediction evaluated at the MAP estimate (black solid line), the 95% credible interval (dark gray), and the 95% prediction interval (light gray). Subplots are organized as follows: (a) S1, (b) S2, (c) S3, and (d) S4.

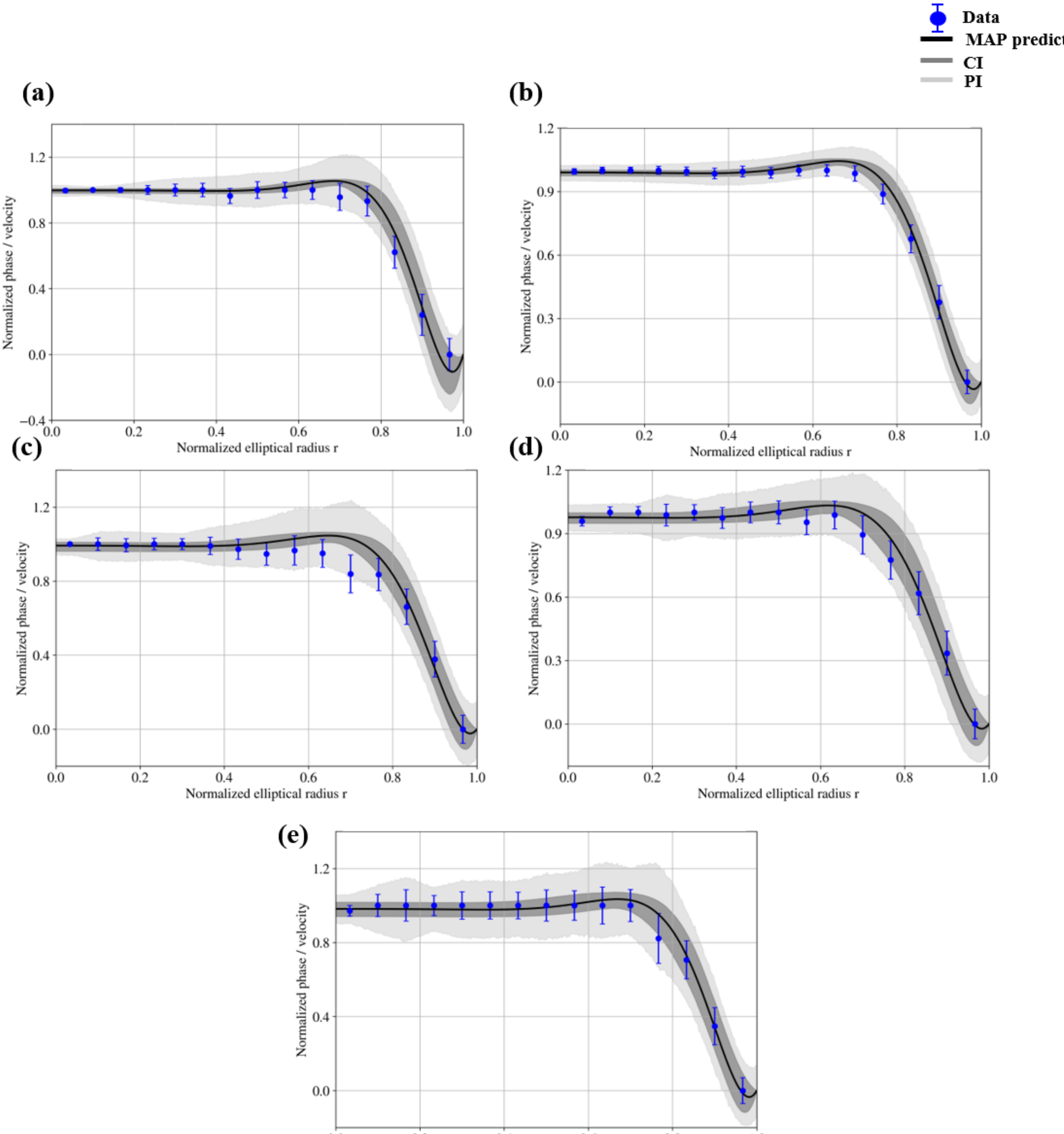

**Figure 5.** Womersley fitting results for all canine subjects. The figure shows the binned radial velocity data with error bars (blue), along with the posterior model prediction evaluated at the MAP estimate (black solid line), the 95% credible interval (dark gray), and the 95% prediction interval (light gray). Subplots are organized as follows: (a) C1, (b) C2, (c) C3, (d) C4, and (e) C5.

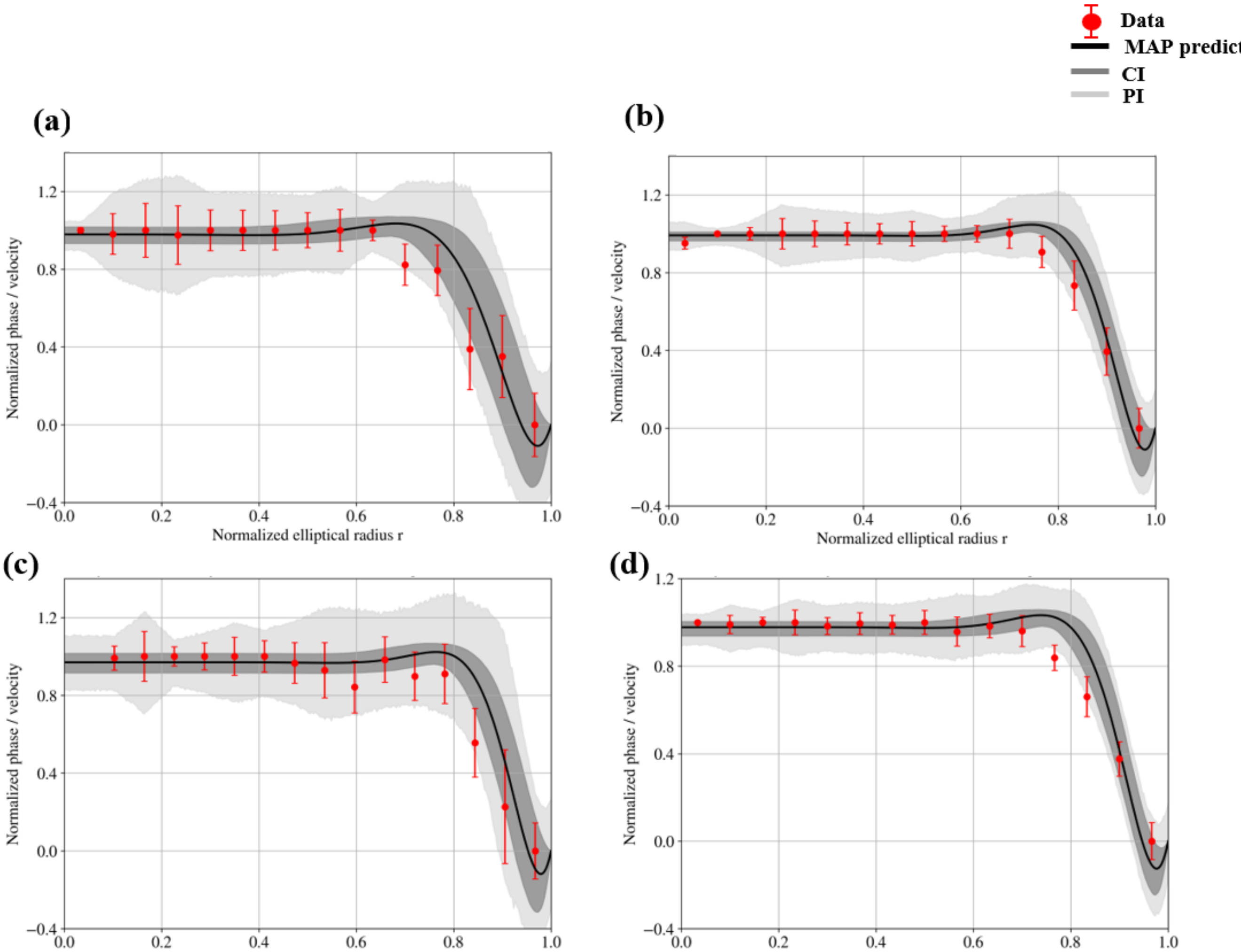

**Figure 6.** Womersley fitting results for all swine subjects. The figure shows the binned radial velocity data with error bars (red), along with the posterior model prediction evaluated at the MAP estimate (black solid line), the 95% credible interval (dark gray), and the 95% prediction interval (light gray). Subplots are organized as follows: (a) S1, (b) S2, (c) S3, and (d) S4.

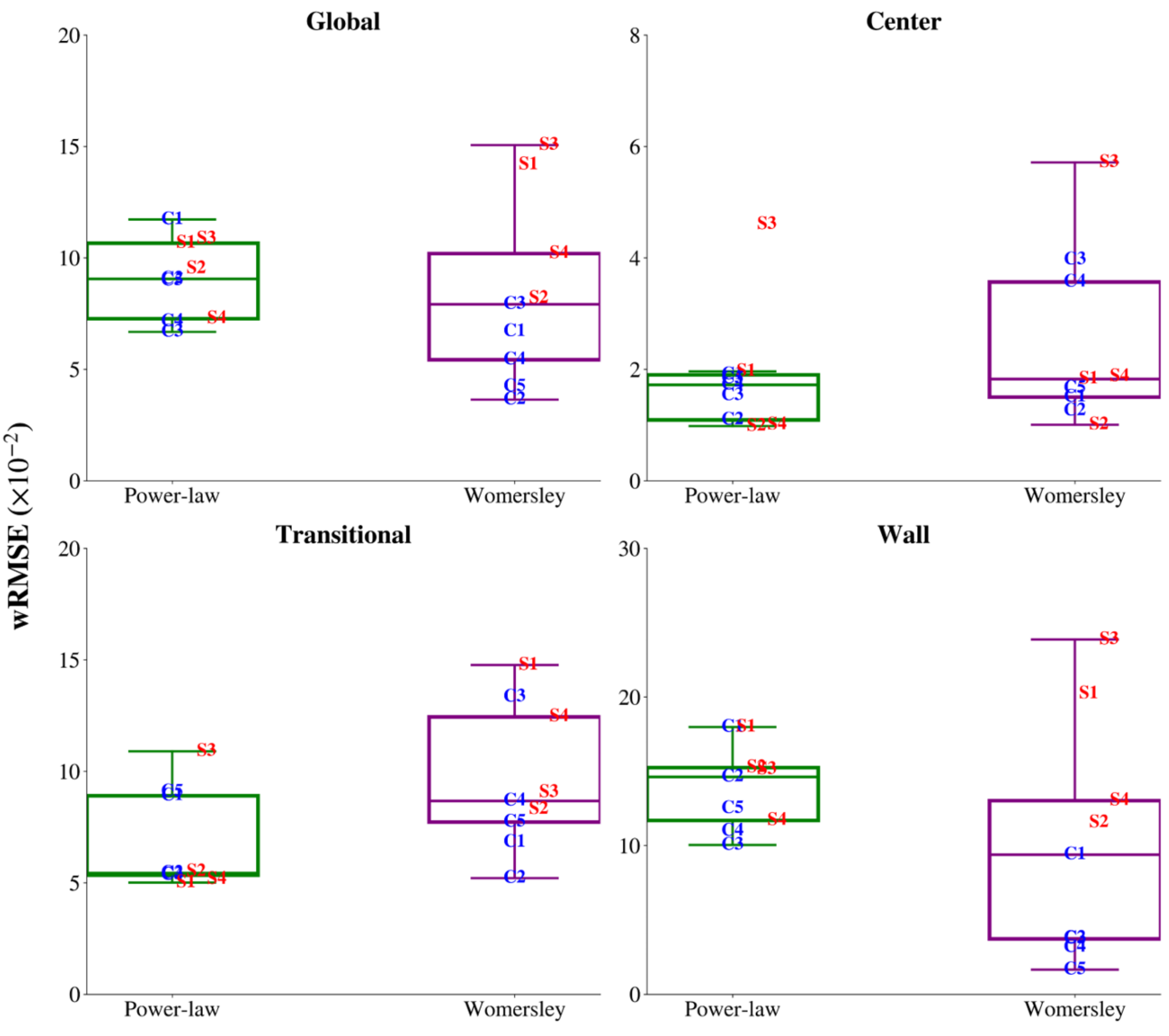

**Figure 7.** Box-and-whisker plots of wRMSE comparing the Power-law (green) and Womersley (purple) models across different vessel regions. Panels show (a) global/regional wRMSE, (b) MPA center region, (c) transitional region, and (d) MPA near-wall region. Blue markers represent canine subjects, while red markers represent swine subjects.

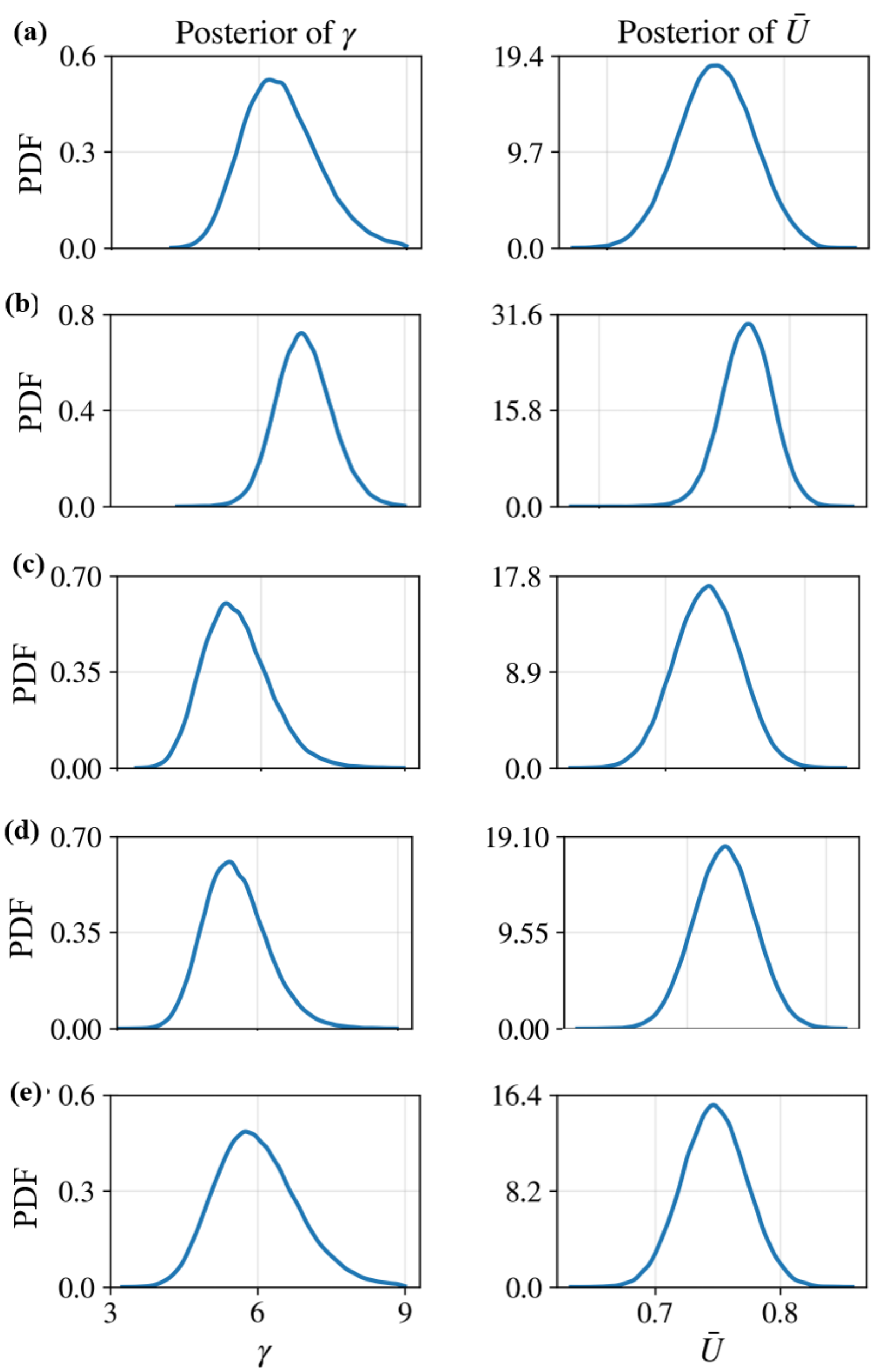

**Figure 8** Marginal posterior distributions for the Power-law model parameters in canine subjects. The left y-axis represents the informative prior PDF, while the right y-axis represents the flat prior PDF. Subplots are organized as follows: (a) C1, (b) C2, (c) C3, (d) C4, and (e) C5.

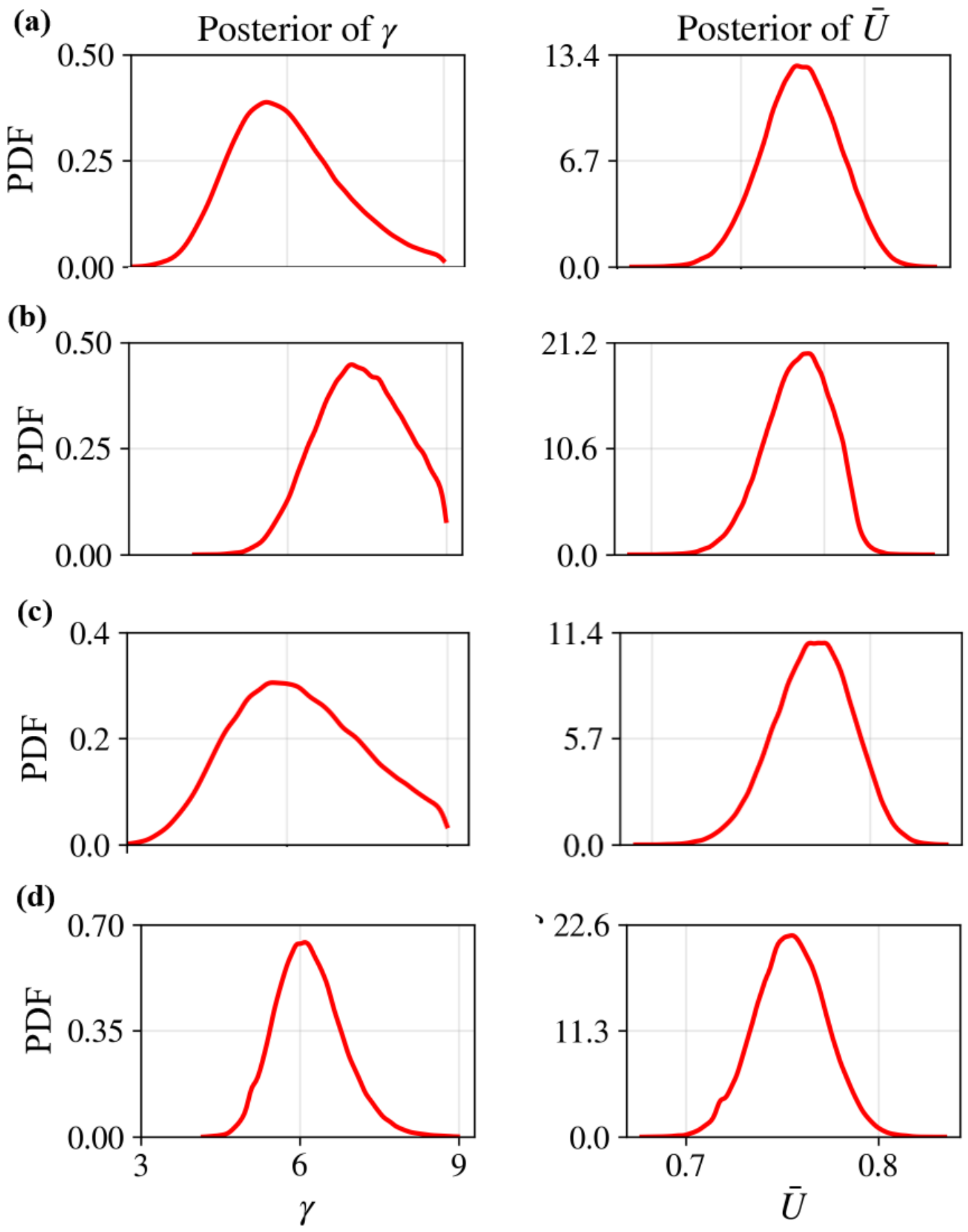

**Figure 9** Marginal posterior distributions for the Power-law model parameters in swine subjects. The left y-axis represents the informative prior PDF, while the right y-axis represents the flat prior PDF. Subplots are organized as follows: (a) S1, (b) S2, (c) S3, and (d) S4.

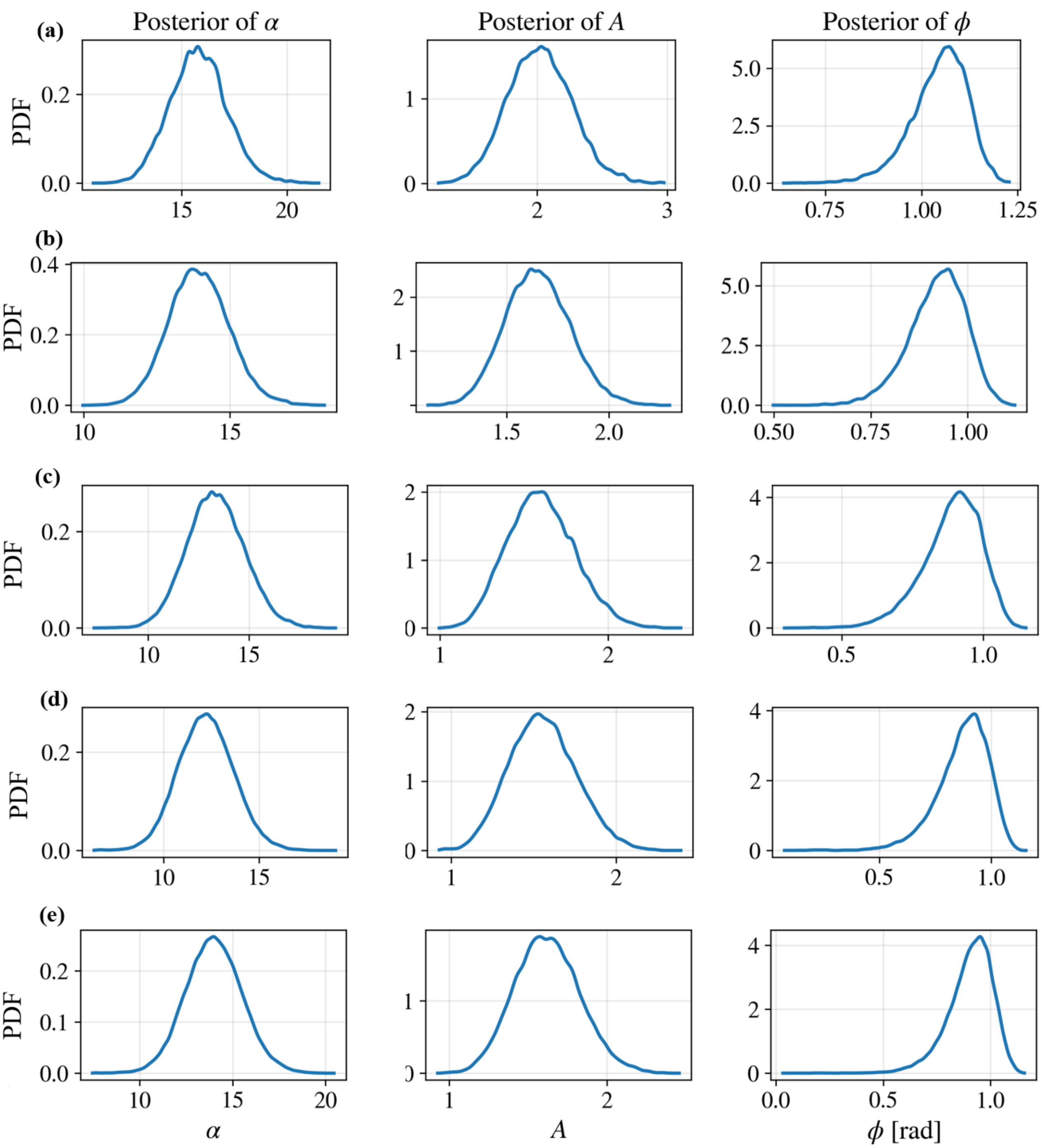

**Figure 10** Marginal posterior distributions for the Womersley model parameters in canine subjects. The left y-axis represents the informative prior PDF, while the right y-axis represents the flat prior PDF. Subplots are organized as follows: (a) C1, (b) C2, (c) C3, (d) C4, and (e) C5.

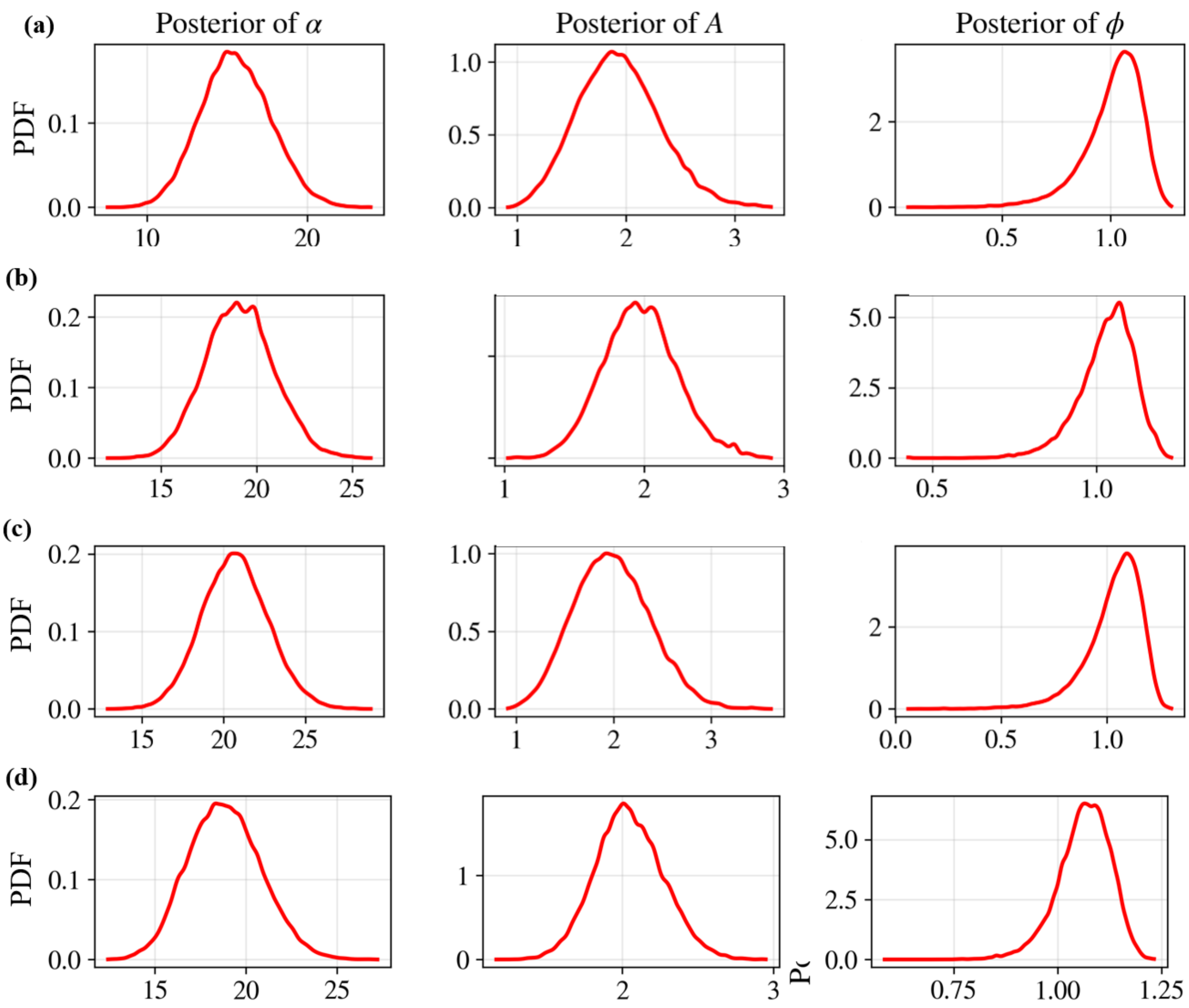

**Figure 11** Marginal posterior distributions for the Womersley model parameters in swine subjects. The left y-axis represents the informative prior PDF, while the right y-axis represents the flat prior PDF. Subplots are organized as follows: (a) S1, (b) S2, (c) S3, and (d) S4.